\documentclass[jkps,fleqn,twocolumn,showpacs,showkeys,floatfix,superscriptaddress]{revtex4}
\usepackage{graphicx,mathrsfs,amssymb}

\begin{document}

\def\br{{\bf r}}
\def\post{{\mbox{\scriptsize post}}}
\def\pre{{\mbox{\scriptsize pre}}}

\title{Two symmetry breaking mechanisms for the development \\
  of orientation selectivity in a neural system}
\author{Myoung Won \surname{Cho}}
\email{mwcho@sungshin.ac.kr}
\affiliation{Department of Global Medical Science,
  Sungshin Women's University, Seoul 142-732}
\date{\today}

\begin{abstract}
Orientation selectivity is a remarkable feature of the neurons located in the
primary visual cortex.
Provided that the visual neurons acquire orientation selectivity through
activity-dependent Hebbian learning, the development process could be
understood as a kind of symmetry breaking phenomenon in the view of physics.
The key mechanisms of the development process are examined here in a neural
system.
Found is that there are at least two different mechanisms which lead to the
development of orientation selectivity through breaking the radial symmetry in
receptive fields.
The first, a simultaneous symmetry breaking mechanism, bases on the competition
between neighboring neurons, and the second, a spontaneous one, bases on the
nonlinearity in interactions.
It turns out that only the second mechanism leads to the formation of a
columnar pattern which characteristics accord with those observed in an animal
experiment.
\end{abstract}


\maketitle


\section{Introduction}
The primary visual cortex (V1) is the first cortex area which receives visual
signals from the retina via the lateral geniculate nucleus (LGN).
The V1 neurons have the typical property to make distinct responses to a small
set of visual stimuli\cite{Hubel1962}.
They have individual receptive fields on the retina, where the center of the
receptive fields becomes the primary feature of the V1 neurons.
The next important feature is the orientation selectivity, through which the V1
neurons detect local bars or edges at the early stage of visual processing.
The neuronal responses also discriminate small changes in spatial frequencies
and colors.
Furthermore, the V1 neurons have ocular dominance (OD), namely turning to one of
the two eyes.
It is not that the V1 neurons have individual features randomly.
They cluster together with others exhibiting similar features and form a
columnar pattern structure, or the called self-organizing feature
map\cite{Kohonen1984}.
The distribution of receptive field centers forms the retinotopic map.
And, a columnar pattern of orientation selectivity (or OD) becomes the
orientation preference (OP) (or OD) map.

A number of models have been suggested to unravel the mechanism of visual map
formation\cite{Erwin1995,Swindale1996}.
Some of them adopt low-dimensional vectors to represent the features of the V1
neurons\cite{Durbin1990,Obermayer1992,Hoffsummer1995,Scherf1999,Goodhill2000,
Cho2004,Cho2005}.
For example, the features of V1 neuron $i$ can be expressed by the vector with
five components $(x_i,y_i,p_i\cos2\phi_i,p_i\sin2\phi_i,z_i)$, where
$(x_i,y_i)$ and $(p_i\cos2\phi_i,p_i\sin2\phi_i)$ represent the center of
receptive field (or retinotopy) and the orientation preference, respectively.
And, $z_i$ represents the ocular dominance.
The abstract models provide with the advantageousness in simulation and
analysis of visual map formation.
They help to reproduce or explain the typical characteristics of columnar patterns
observed in V1 easily.
Interestingly, ignoring retinotopy, the low-dimensional vector corresponds with
a spin model vector\cite{Tanaka1989,Cowan1991}.
Moreover, the many characteristics of an OP or OD map have analogy with those
of a magnetic system\cite{Cho2004,Cho2005}.
For example, an OP map embraces singular points, dubbed pinwheels, around which
preferred angles change by multiples of 180$^\circ$ along a closed loop.
Pinwheels corresponds to (in-plane) vortices in magnetism, and are used to be
annihilated during the visual map formation as vortices
do\cite{Wolf1998,Kosterlitz1973}.

Meanwhile, some other models adopt high-dimensional feature vectors, which
components represent the afferent connectivity from the retina ganglion cells
(RGCs) (or the LGN cells) to the V1 neurons in usual.
A high-dimensional feature representation model, reflecting more biological
factors, is essential for inquiring how the V1 neurons acquire their typical
features through a learning process.
A correlation-based learning model, adopting high-dimensional feature vectors,
was suggested to explain the formation of the OP or the OD map\cite{Miller1989,
Miller1992,Miller1994}.
And, it was introduced a simplified version of the correlation-based learning
model, which adopts only linear interaction terms\cite{Dayan2001}.
It was also demonstrated that such a linear learning model can lead to the
formation of a retinotopic (or topographic) map\cite{Cho2012}.


Analytically, the development of a typical feature in a neural system relates
to the existence of a symmetric property in external inputs.
It is the symmetry in left and right RGC activities which becomes the origin of
the OD development\cite{Dayan2001}.
And, the translational symmetry in RGC activities becomes the cause of the
retinotopic map regulation\cite{Cho2012}.
Note that the organization of a columnar pattern relates to the translational
symmetry in lateral interactions.


Analogically, the rotational symmetry in visual inputs may relate to the
development of orientation selectivity; however, there are different opinions
about the detail mechanism.
The discharges of RGCs are determined by the convolution visual inputs with
center-surround receptive fields, where the center is either ON or OFF while
the surround is the opposite.
A major suggestion is that the two different types of receptive field is the
essence of the orientation selectivity development:
There are the models which explain the feature development by the competition
or the Moire\'{e} interference between ON- and OFF-center type
RGCs\cite{Miller1992,Miller1994,Paik2011}.
On the other hands, there ia the suggestion that the development of orientation
selectivity is possible independently of OFF-center RGCs\cite{Cho2009}.
It is known that the receptive field of neurons have a near Gaussian
distribution shape when a topographic map develops in a neural network with
homogeneous input neurons\cite{Cho2012}.
The neurons would acquire the feature of orientation selectivity additionally
if they lost the radial symmetry in the receptive fields,


Based on the view, the development of orientation selectivity is investigated
here in a learning model.
The model, derived from biological neural dynamics and synaptic plasticity, has
an extended form of the linear learning model which was studied for the
topographic map formation.
Observed is that there are at least two different mechanisms which lead to the
symmetry breaking in receptive fields for the development of orientation
selectivity.
The first is a simultaneous symmetry breaking mechanism, bases on competitive
interactions between neurons: Under a large degree of competition, receptive
fields should be squeezed in order to reduce the overlap with neighbors.
While the first mechanism is regardless of nonlinear interactions, the second
requires a high order interaction term for the development of orientation
selectivity.
The second is a spontaneous symmetry breaking mechanism, through which the
radial symmetry of receptive fields are broken independently of interactions
with neighbors.
Both the mechanisms lead to the development of an OP columnar pattern with
singular points; however, the emergent columnar pattern has the characteristics
being somewhat different (or coincide) with those in experimentally observed
ones when the first (or the second) mechanism leads the symmetry breaking
phenomenon.



\section{The model}
Suppose a neural network composed of input-output two layers, where output
neurons have flexible feedforward (unidirectional) connections from input
neurons and static lateral (bidirectional) connections with other output
neurons.
Suppose the continuous function $\phi_\ell(t)$ $(\in[0,1])$ stands for the
firing probability of neuron $\ell$ at position $\br_\ell=(x_\ell,y_\ell)$ and
time $t$ averaged over trials.
Labeling input and output neurons with indices $(a,b)$ and $(i,j)$,
respectively, suppose the firing probability of neurons are expressed by the
equation
\begin{eqnarray}
  \phi_a(t)=\eta_a(t)+h_a(t)
\end{eqnarray}
and
\begin{eqnarray} \label{eq:phi}
  \phi_i(t)=\eta_i(t)+\sum_a\big[D_{ia}\phi_a\big](t)
    +\sum_{j,a}\big[D_{ij}D_{ja}\phi_a\big](t) \\
  +\sum_{a,b}\big[D_{iab}^{(3)}\phi_a\phi_b\big](t). \hspace{2.0cm} \nonumber
\end{eqnarray}
Here $\eta_\ell(t)$, assumed to be generated by Poisson process, represent
endogenous neural firings due to noisy currents and $h_a(t)$ neural firings due
to external visual inputs.
With the brief notation
$[D_{i\ell}\phi_\ell](t)\equiv\int dsD_{i\ell}(s)\phi_\ell(t{-}s)$ and
$[D_{iab}^{(3)}\phi_a\phi_b](t)\equiv\int dsds'D_{iab}^{(3)}(s,s')
\phi_a(t{-}s)\phi_b(t{-}s')$, the interaction strengths are expressed in the
form
\begin{eqnarray}
  D_{i\ell}(s)=\lambda(s)W_{i\ell}
\end{eqnarray}
and
\begin{eqnarray}
  D_{iab}^{(3)}(s,s')=\lambda^{(3)}(s,s')W_{ia}W_{ib}.
\end{eqnarray}
Here $\lambda(s)$ and $\lambda^{(3)}(s,s')$ relate to the time delay in firing
propagation, and $W_{i\ell}$ are the scaled coupling strengths.
Experimentally, $\lambda(s)$ and $\lambda^{(3)}(s,s')$ could be measured by the
derivative of the firing propagation probability with the respect to the scaled
coupling strengths\cite{Cho2014}.



Meanwhile, a biological synapse changes its efficacy depending on the precise
difference between post- and presynaptic firing times\cite{Markram1997,Bi1998}.
According to the mechanism, the called spike-timing-dependent plasticity (STDP)
rule, the rate of change in feedforward connection strength can be expressed in
the from
\begin{eqnarray} \label{eq:STDP}
  \Delta W_{ia}=\int dt\int dt'\,F(t{-}t')\phi_i(t)\phi_a(t').
\end{eqnarray}
Here $F(t)$ describes the dependence of the synaptic modification on the
difference between the post- and the presynaptic spike times.
Usually $F(t)$ produces long-term potentiation (LTP) or positive change
in a synapse when $t>0$, and long-term depression (LTD) or negative change
otherwise.


Substitution of Eq.(\ref{eq:phi}) into Eq.(\ref{eq:STDP}) leads to the learning
model for the visual map formation expressed as
\begin{eqnarray} \label{eq:DW}
  \Delta W_{ia}=\sum_jJ_{ij,a}W_{ja}+\sum_bW_{ib}C_{ba}
    +\sum_{b,c}W_{ib}W_{ic}V_{bca},
\end{eqnarray}
where
\begin{eqnarray}
  J_{ij,a}&=&W_{ij}\sum_a\int dt\int dt'\,F(t{-}t')
    \big[\lambda\lambda\phi_a\big](t)\phi_a(t') \\
  C_{ba}&=&\int dt\int dt'\,F(t{-}t')\big[\lambda\phi_b\big](t)\phi_a(t') \\
  V_{bca}&=&\int dt\int dt'\,F(t{-}t')\big[\lambda^{(3)}\phi_b\phi_c\big](t)
    \phi_a(t')
\end{eqnarray}

First, the expectation value $J_{ij,a}/W_{ij}$ produce a similar value for all
$a$ when the activity of input neurons are similar, and become a positive value
when $F(t)>0$ for $t>0$ because $\lambda(t)$ has positive values only for small
positive values of $t$ in usual.
Considered the lateral connections $W_{ij}$ become positive in short-range and
negative in long-range in a cortex area, the lateral interactions is expressed 
in the form
\begin{eqnarray} \label{eq:J_ij}
  J_{ij,a}=J_{ij}\equiv\varepsilon_{ij}-\gamma,
\end{eqnarray}
where $\varepsilon_{ij}$ equal unity for a nearest-neighbor pair $(i,j)$, and
vanish otherwise.
It is also possible to derive the negative part of the lateral interactions
independently of inhibitory connections if the firing propagations and the form
of $F(t)$ are considered in detail\cite{Cho2009,Cho2012,Cho2014}.


Meanwhile, $C_{ab}$ is determined by the firing correlation of between two
input neurons.
If $\int dt\,F(t)<0$ and $F(t)>0$ for $t>0$, the expectation value changes its
sign from positive to negative with decreasing input correlation, where the
correlation decreases with the distance between input neurons in
usual\cite{Cho2014}.
Based on the property, the input correlation matrix is expressed in the from
\begin{eqnarray} \label{eq:C_ab}
  C_{ab}=\alpha\exp\left[-\frac{\Lambda_{ab}}{\sigma^2}\right]-\beta,
\end{eqnarray}
where $\Lambda_{ab}$ is set by $|\br_a{-}\br_b|^2/2$.

Similarly, the tensor $V_{abc}$ is determined by the firing correlation of
three input neurons, which has the dependency on not only the distance between
but also the linearity of neural positions because of frequent edge patterns in
natural images.
On that account, the input correlation tensor is expressed in the form
\begin{eqnarray} \label{eq:V_abc}
  V_{abc}=\alpha'\exp\left[-\frac{3\Lambda_{abc}^+}{2\sigma_1^2}
    -\frac{3\Lambda_{abc}^-}{2\sigma_2^2}\right]-\beta'.
\end{eqnarray}
Here $\displaystyle\Lambda_{abc}^\pm=\mbox{tr}(\Sigma)\pm\sqrt{[\mbox{tr}
(\Sigma)]^2{-}4\det(\Sigma)}$ are the eigenvalues of the $2{\times}2$
covariance matrix $\Sigma=E\big[(\br-\bar{\br})(\br-\bar{\br})^\top\big]$ with
$\bar{\br}=E[\br]=(1/3)\sum_{\mu=a,b,c}\br_\mu$, or the variance along the
major and minor axis in the multivariate Gaussian distribution of $\br_a$,
$\br_b$, $\br_c$ three positions.
Note that $\Lambda_{ab}$ in Eq.~(\ref{eq:C_ab}) also becomes the variance in
the distribution of $\br_a$, $\br_b$ two positions.

In addition, it is assumed that the length of the feedforward connection vector
$|{\bf W}_i|^2=\sum_aW_{ia}^2$ for all $i$ is normalized to unity during the
evolution.


\begin{figure}[t]
\includegraphics[width=8cm]{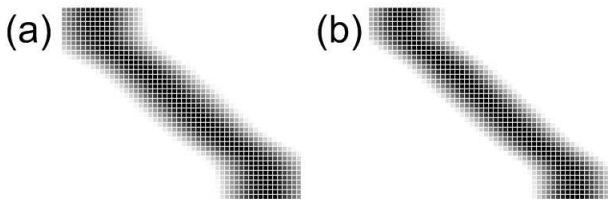}
\caption{ \label{fig:1D}
Learning results in a network composed of two layers with one-dimensional
structure.
Interaction parameters are given by (a)
$(\alpha,\beta,\sigma,\alpha',\beta')=(0.3,0.05,16,0,0)$ and (b)
$(\alpha,\beta,\alpha',\beta',\sigma_1)=(0,0,0.3,0.05,16)$, and $\gamma=0.02$
and $\Lambda_{abc}^-=0$ for all cases.
The network size is $N=40$ and $M=50$ for output and input layers, respectively.
}
\end{figure}



\section{Results}
It is possible to predict from the previous studies what features would emerge
when the learning rule has no nonlinear term.
Reported was that a linear model with the form $\Delta W_{ia}{=}[JW{+}WC]_{ia}$
could lead to the development of a topographic map when both $J$ and $C$
produce positive values for close neurons and negative values
otherwise\cite{Cho2012}.
Especially, the solution of the linear model is well studied for a neural
network with a one-dimensional lattice structure:
Close output neurons would have connections from close input neurons, where
in-coming connections into the output neurons and out-going connections from
the input neurons form a near Gaussian distribution.
Here the distribution of in-coming connections from input neurons to an output
neuron becomes the receptive field of the output neuron.
However the nonlinear term is considered, the correlation tensor $V$ should
exert the same effects on the network formation with the correlation matrix $C$
because the three-point correlations contain no more typical feature for the
one-dimensional lattice structure.
Figure \ref{fig:1D}(a) (or (b)) show the emergent connection structure in a
neural network when only $J$ and $C$ (or $V$) provides with valid coefficients.

Meanwhile, the learning model has the possibility to have the other type
solutions when the neural network has a two-dimensional lattice structure.
The Gaussian distribution form of receptive fields could polarized when they
have difference variances along major and minor axes.
And then, it deserves that the neurons acquire orientation selectivity through
breaking of the radial symmetry in the receptive field.

\begin{figure}[t]
\includegraphics[width=7.5cm]{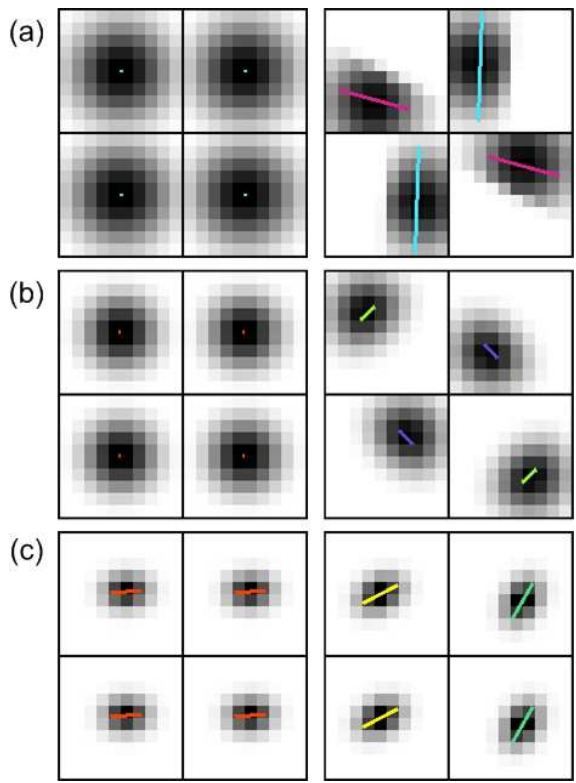}
\caption{ \label{fig:2x2}
Emergent receptive fields in a networks composed of two layers with square
lattice structure.
The lines represent the direction of and the degree of orientation selectivity.
Interaction parameters are given by (a)
$(\alpha,\beta,\sigma,\alpha',\beta')=(0.1,0.03,20,0,0)$ and (b-c)
$(\alpha,\beta,\alpha',\beta',\sigma_1)=(0,0,0.1,0.02,20)$ with $\sigma_2=5$
and $0.4$, respectively, in the network of size
$(N,M)=(2{\times}2,10{\times}10)$.
The parameter $\gamma$ is given by $1$ in the left figures, and $4$, $4$, $2$,
respectively, in the right figures.
}
\end{figure}
\begin{figure}[t]
\includegraphics[width=8.4cm]{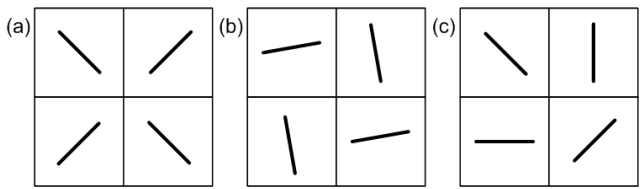}
\caption{ \label{fig:pinwheel}
Illustration of singularities, around which the orientation changes by
multiples of (a-b) 360$^\circ$ or (c) 180$^\circ$ along a closed loop.
}
\end{figure}

In order to investigate what factor causes such symmetry breaking, it is first
examined the effects of the parameter $\gamma$, determining the degree of
competition between neurons, on the network formation.
It is known that there is a critical value in the degree of competition only
above which localized receptive fields develop\cite{Cho2012}.
If the degree of competition becomes more severe, the form of receptive fields
would be distorted to reduce the overlap between neighbors.

Figure \ref{fig:2x2}(a) shows the emergent receptive fields for different
$\gamma$, where the effects of the tensor $V$ is ignored by setting
$(\alpha',\beta')=(0,0)$ so that the network formation is led by only the
linear interaction terms.
In comparison with the near isotropic shape for an intermediate value of
$\gamma$ in the left figure, the form of receptive fields become elliptical for
a large value of $\gamma$ in the right figure.
A similar phenomenon is observed when the network formation is led by not $C$
but $V$ (Fig.~\ref{fig:2x2}(b)).
Namely, the symmetry breaking mechanism is regardless of the nonlinearity in
the learning model.


Nevertheless, the emergent columnar patterns in Fig.~\ref{fig:2x2}(a) and (b)
have something difference characteristics with experimentally observed ones.
They possess a single singular point, around which the preferred angle changes
by 360$^\circ$ along a closed loop; however, pinwheels in an OP map are
singular points around which the preferred angle changes by 180$^\circ$ (see
Fig.~\ref{fig:pinwheel}).
In addition, Although pinwheels, understood as a topologically excited state,
are used to disappear during a cortical development process, the singular
points do not vanish to the end in the simulations\cite{Wolf1998,Cho2004}.
The phenomenon is caused by that the singular points develop inevitably when a
group of neurons acquire orientation selectivity simultaneously through the
squeeze of neighboring receptive fields.


\begin{figure}[t]
\includegraphics[width=7.6cm]{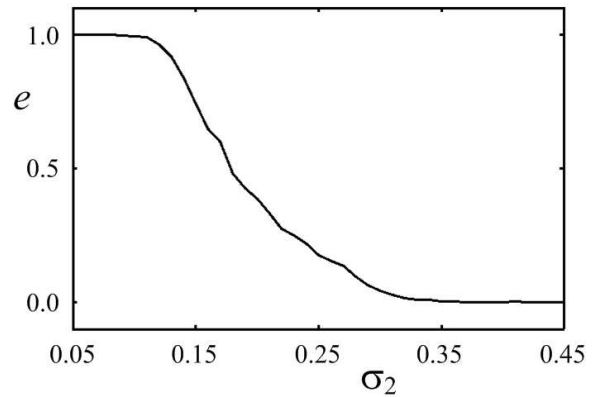}
\caption{ \label{fig:eccentricity}
Change in the eccentricity of receptive fields, depending on $\sigma_2$.
Interaction parameters 
$(\alpha,\beta,\alpha',\beta',\sigma_1,\gamma)=(0,0,0.1,0.01,20,0)$ have been
used in a network of size $(N,M)=(1{\times}1,8{\times}8)$.
}
\end{figure}


Next, the emergent shape of receptive fields is examined when the correlation
tensor $V$ has a high degree of anisotropy, where the degree is controlled by
the ratio of $\sigma_2$ to $\sigma_1$.
Figure \ref{fig:2x2}(c) shows that orientation selectivity develops when
$\sigma_2$ is much smaller than $\sigma_1$ and the effect of the correlation
matrix $C$ is ignored by setting $(\alpha,\beta)=(0,0)$.
Observed is that the neurons acquire orientation selectivity independently of
the degree of competition, but form different columnar patterns depending on
the degrees.
They have similar preferred angles with neighbors for a small $\gamma$ in the
left figure and different ones for a large $\gamma$ in the right figure.
The effect of competition on the columnar pattern accords with the prediction
in the most low-dimensional feature representation models:
The degree of competition determines the gradient when the preferred angles
change smoothly along a direction in a patch of the columnar
pattern\cite{Obermayer1992,Durbin1990,Scherf1999,Goodhill2000,Hoffsummer1995,
Cho2004,Cho2005}.

Compared with the first mechanism, the second deserves to be a spontaneous
symmetry breaking mechanism because a single neuron could acquire orientation
selectivity independently of the existence of or the interaction with
neighboring neurons.
In addition, the existence of a singular point is not indispensable to the
emergence of orientation selectivity, as shown in Fig.~\ref{fig:2x2}(c).
It is also investigated the change in the eccentricity of receptive fields,
depending on $\sigma_2$ for a fixed $\sigma_1$, in a neuron system with a
single output neuron (Fig.~\ref{fig:eccentricity}).
The graph exhibits that the isolated neuron could have an elliptical receptive
field when $\sigma_2$ is larger than a critical value, and the eccentricity
decreases with increasing $\sigma_2$.

\begin{figure}[t]
\includegraphics[width=7cm]{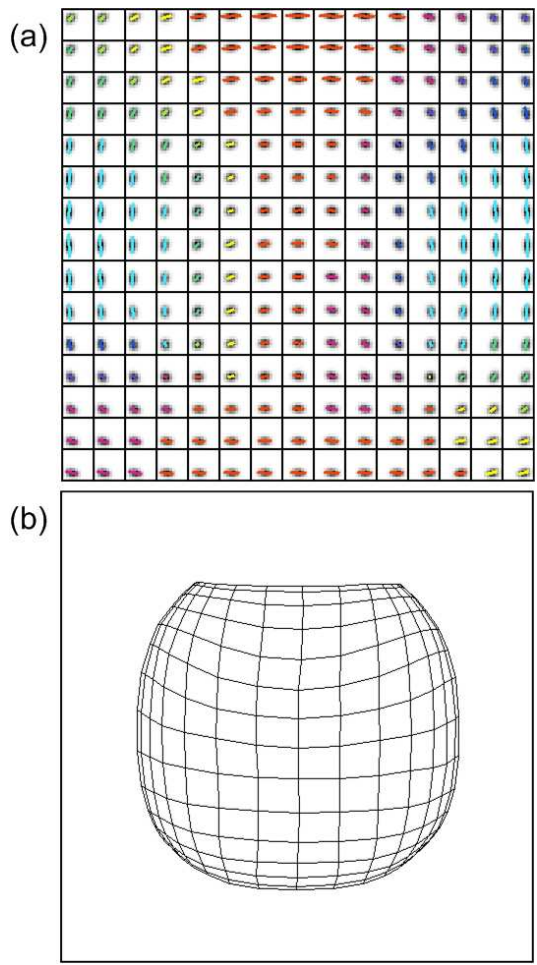}
\caption{ \label{fig:OP}
Distribution of (a) orientation selectivity and (b) topography after learning
in a network composed of two layers with square lattice structure.
Interaction parameters 
$(\alpha,\beta,\alpha',\beta',\sigma_1,\sigma_2,\gamma)
=(0,0,0.5,0.1,20,0.4,1)$ have been used in a network of size
$(N,M)=(15{\times}15,10{\times}10)$.
}
\end{figure}

Finally, it is examined the development of a columnar pattern with the
orientation selectivity as well as the topography in a larger lattice structure
(Fig.~\ref{fig:OP}).
The center of receptive fields form a proper topographic map, and the polarity
of them an OP map. 
The OP map have several singular points around which the preferred angles
change by multiples of 180$^\circ$.

\section{Discussion}
In this paper, it has been examined in a learning model that the development of
orientation selectivity and the formation of columnar patterns.
The model is a much simplified one in consideration of the complex development
process and the detail structure of early visual systems, such as the genetic
effects, the relay of visual signals via LGN, the different types of RGCs, and
so on; however, such a minimal model is helpful to understand the essential
mechanisms for the development of typical properties in a neural network.

In the case of retinotopy or OD, it is known that a linear model can lead to
the development of the neural feature and the formation of a proper columnar
pattern.
The key mechanism of both the feature development is the (block)
diagonalization of the input correlation matrix depending on the symmetry in
visual inputs\cite{Dayan2001,Cho2012}.
Analogically, it was also suggested that the development of orientation
selectivity may be explained similarly by the diagonalization of the
correlation matrix via the rotational symmetry in visual inputs\cite{Cho2009}.
Nevertheless, it has been not demonstrated until now whether the development of
a proper OP map is possible just by adjusting the form of the correlation
matrix in a linear model.

Observed in this paper is that the development of orientation selectivity is
possible in a linear model; however, the feature development mechanism does not
relate to the correlation matrix diagonalization.
The mechanism bases on the competitive relationship between neighbor neurons,
where the degree is determined not in the input correlation but in the lateral
interaction matrix.
Nevertheless, it is difficult to explain the OP development in the brain by the
the first mechanism.
However the mechanism leads to the development of a columnar pattern with
different preferred angles and singular points; the emergent pattern has
different characteristics with experimentally observed ones.

Meanwhile, the other mechanism of orientation selectivity development is found
when the model is extended to have a high order interaction term.
The requisite of the nonlinearity for the model seems to be natural in
consideration of that the effect of frequent lengthy patterns in visual
scenes can be reflected only by the input correlation between three or more
neurons.
The characteristics of an emergent columnar pattern also coincide with those in
the experimentally observed ones.
The second is a kind of spontaneous symmetry breaking mechanism, i.e., a single
neuron could acquire orientation selectivity independently of the interactions
with neighbors.
The spontaneousness is in close connection with the instability of pinwheels,
i.e., individual receptive fields should have orientation selectivity
independently, like spins do, if pinwheels are indeed topologically excited
states.
The second mechanism is also an interesting one in the view of physics if it
is considered that there are several models to explain a spontaneous symmetry
breaking phenomenon by the effect of a higher-order interaction
term\cite{Goldstone1961}.

\bibliography{op}

\end{document}